\documentclass[useAMS,usenatbib]{mn2e}\usepackage{epsfig}\usepackage{amsmath}

\newcommand{\be}{\begin{equation}}
\newcommand{\beq}{\begin{equation}}
\newcommand{\ba}{\begin{eqnarray}}
\newcommand{\ee}{\end{equation}}
\newcommand{\eeq}{\end{equation}}
\newcommand{\ea}{\end{eqnarray}}

\def\lsim{~\rlap{$<$}{\lower 1.0ex\hbox{$\sim$}}}

\def\gtrsim{~\rlap{$>$}{\lower 1.0ex\hbox{$\sim$}}}

\title[UV-faint Galaxies at $z\ga4$]{A Predicted New Population of UV-faint Galaxies at $z\ga4$}

\author[Wyithe, Loeb \& Oesch]{J. Stuart B. Wyithe$^{1}$, Abraham Loeb$^{2}$, Pascal A. Oesch$^{3,4}$\\$^1$
School of Physics, University of Melbourne, Parkville, Victoria,
Australia\\$^2$ Astronomy Department, Harvard University, 60 Garden
Street, Cambridge, MA 02138, USA\\$^3$ UCO/Lick Observatory, University of California, Santa Cruz, CA 95064, USA\\$^4$ Hubble Fellow\\Email: swyithe@unimelb.edu.au; aloeb@cfa.harvard.edu; poesch@ucolick.org}

\begin{document}


\maketitle

\label{firstpage}
\begin{abstract}

We show that a bursty model of high redshift star formation explains several puzzling observations of the high redshift galaxy population. We begin by pointing out that the observed specific star formation rate requires a duty-cycle of $\sim10\%$, which is much lower than found in many hydro-dynamical simulations. This value follows directly from the fact that the observed star formation rate in galaxies integrated over a Hubble time exceeds the observed stellar mass by an order of magnitude.  We use the large observed specific star formation rate to calibrate the efficiency of feedback in a model for the high redshift star formation rate which includes merger driven star formation regulated by SNe feedback. This model reproduces the star formation rate density function and the stellar mass function of galaxies at $4\la z\la7$. A prediction of the model is that the specific star formation rate does not evolve very rapidly with either mass or redshift, in agreement with observation. This is in contrast to results from hydrodynamical simulations where the star formation closely follows the accretion rate, and so increases strongly towards high redshift. The bursty star formation model naturally explains the observation that there is not enough stellar mass at $z\sim2-4$ to account for all of the star-formation observed, without invoking properties like an evolving initial mass function of stars. The finding of a duty cycle that is $\sim10\%$ implies that there should be ten times the number of known galaxies at fixed stellar mass that have not yet been detected through standard UV selection at high redshift. We therefore predict the existence of a large undetected population of UV-faint galaxies that accounts for most of the stellar mass density at $z\sim4-8$.

\noindent 

\end{abstract}


\section{Introduction}

The galaxy luminosity function is the primary observable that must be
reproduced by any successful model of galaxy formation. At
$z\gtrsim6$, it also represents one of the most important observables
for studying the reionization of cosmic hydrogen.  The luminosity function of Lyman-break galaxy (LBG) candidates discovered at
$z\ga6$ in recent WFC3/IR surveys is described by a Schechter
function with characteristic density $\Psi_\star$ in comoving
Mpc$^{-3}$, and a power-law slope $\alpha$ at luminosities $L$ below a
characteristic break $L_\star$
\citep[e.g.][]{Bouwens2011,Oesch2012,Loeb2012,McLure2013,Ellis2013,Oesch2013b}. Developing a
theoretical picture of the important processes involved in setting the
star formation rate at high redshift lies at the forefront of
understanding this important cosmic epoch
\citep[e.g.][]{Trenti2010,Finlator2011,Munoz2011,Raicevic2011,Salvaterra2011}.

Complex hydrodynamical models have been used to model the observed
properties of high redshift galaxies. For
example \citet[][]{Finlator2011} have modelled the growth of stellar
mass in high redshift galaxies using hydrodynamical simulations
coupled with sub-grid models for processes including star formation
and metal enrichment, and broadly reproduce the luminosity function
evolution as well as the blue colours of the young stellar populations
at high redshift.  Similarly, \citet[][]{Salvaterra2011} and
\citet[][]{Jaacks2012} have calculated the evolution of the luminosity
function in detailed numerical simulations including calculations of
enrichment and dust reddening, with the latter also including
additional physics related to the transition from population-III to
population-II stars. While these models are able to reproduce the 
luminosity function and star formation rate density function, they over-produce the high redshift stellar 
mass function, particularly at the low mass end. 

Recently, \citet[][]{Wyithe2013} presented a model for the high redshift star formation rate density function, which includes merger driven star formation regulated by SNe feedback. This model fits a range of observables, and implies a duty cycle of star formation of only 1-10\%, much lower than found in hydro-dynamical simulations. The model successfully predicts the
observed relation between star formation rate and stellar mass at
$z\gtrsim4$. Supernovae feedback was found to lower the efficiency of star
formation in the lowest mass galaxies, making their contribution to
reionization small.

A puzzling observation in recent high redshift galaxy research has been that the star formation rate per stellar mass (specific star formation rate; sSFR) does not seem to evolve significantly with either mass or redshift \citep[e.g.][]{Stark2009,Gonzalez2010}. 
While subsequent analyses indicated that the absolute value of the sSFR at $z\geq4$ might have been underestimated in these first derivations after including updated estimates of dust extinction and accounting for the impact of rest-frame optical emission lines, the current best observational estimates indicate only slow evolution across $z\sim4-7$ \citep[see e.g.][]{Gonzalez2012,Stark2013}. However, the observational debate is far from settled \citep[e.g.][]{Smit2013}.

Most simulations of high redshift galaxy formation do not reproduce the observed plateauing of specific star formation rate at $z>2$. This is because simulations generally associate star formation primarily with the accretion of gas. As a result they predict a rapid increase in the specific star formations rate, which can be understood because the specific accretion rate is found to scale as $(1+z)^{2.5}$ \citep[][]{Neistein2008}. To understand which aspect of high redshift galaxy formation models drives the incorrect prediction of an evolving specific star formation rate, \citet[][]{Weinmann2011} calculated the specific star formation history within a suite of semi-analytic models. At $z>4$, they found that the evolution of specific star formation rate could be reproduced in the presence of strong SNe feedback. In this paper, we find that a SNe regulated model with star-formation triggered by mergers and a low duty cycle naturally reproduces both the large value of specific SFR, and the observed behaviour with mass and redshift.

The relationship between the observed star formation rate and stellar mass per unit volume has also been an observational focus. \citet[][]{Wilkins2008} compiled estimates of stellar mass and star formation rates as a function of redshift in order to investigate whether the integral of star formation rate matches the observed stellar mass. Puzzlingly, at $z\sim2-4$,  \citet[][]{Wilkins2008} find that there is not enough stellar mass to account for all of the star-formation observed. Conversely, at high redshift \citet[][]{Bouwens2011} and \citet[][]{Robertson2013} find that the observed stellar mass is accounted for by the observed star-formation rate \citep[see also][]{Stark2013}. In this paper we argue that both observations can be understood in the context of a star formation model with a duty-cycle of order 10\%.

We begin in \S~\ref{simple} by pointing out the general constraint on the duty-cycle that is provided by observations of the specific star formation rate. Then, in \S~\ref{model} we briefly summarise the model for high redshift star formation presented in \citet[][]{Wyithe2013}.  We next present a comparison of this model with various observables including the star formation rate density function, specific star formation rate, clustering amplitude and stellar mass function in \S~\ref{results}. We discuss the detectability of a predicted population of low UV luminosity galaxies in \S~\ref{lowlum}, and finish with a discussion in \S~\ref{discussion}. In our numerical examples, we adopt the standard
set of cosmological parameters \citep{Komatsu2011}, with values of
$\Omega_{\rm b}=0.04$, $\Omega_{\rm m}=0.24$ and $\Omega_\Lambda=0.76$
for the density parameters of matter, baryon, and dark energy,
respectively, $h=0.73$, for the dimensionless Hubble constant, and
$\sigma_8=0.82$.

\section{The specific star formation rate of star-forming galaxies}
\label{simple}

\begin{figure*}
\begin{center}
\vspace{3mm} \includegraphics[width=15cm]{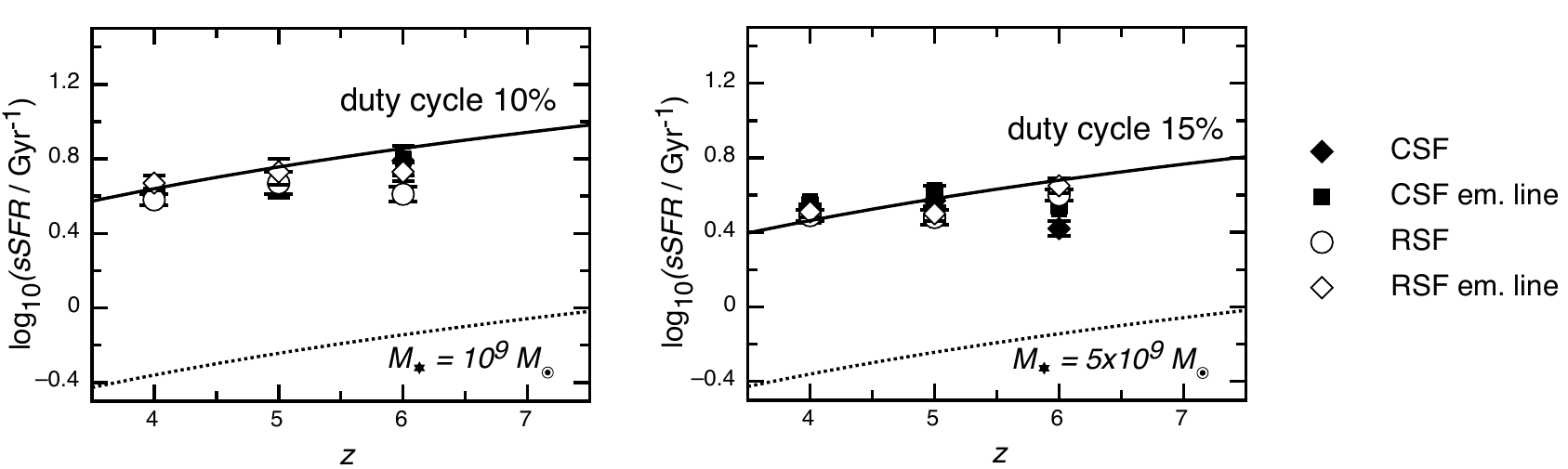}
\caption{\label{fig1} The specific star-formation rate as a
function of redshift calculated based on equation~(\ref{eq:simple}) for duty-cycles of 10\% and 15\% (solid lines) and 100\% (dotted lines), in comparison with measurements at stellar masses of $10^9$ and $5\times10^9$ M$_\odot$. The data points are from \citet[][]{Gonzalez2012}, and the labels represent stellar masses based on the assumptions of a constant (CSF) and rising (RSF) star-formation history without and with emission lines included (em. line) as described in that paper.}
\end{center}
\end{figure*}

Before discussing our particular model for supernovae (SNe) regulated star formation, we begin by looking at the very general constraint on duty-cycle ($\epsilon_{\rm duty,tot}$) provided by the specific star formation rate. In the simplest model of constantly star-forming galaxies, the specific star formation rate is 
\begin{equation}
\label{eq:simple}
sSFR = \frac{SFR}{M_{\star}} = \frac{SFR}{SFR\times(\epsilon_{\rm duty,tot}H^{-1})} = (\epsilon_{\rm duty,tot}H^{-1})^{-1}.
\end{equation}
Thus, the specific star formation rate leads to a direct estimate of the duty cycle of star formation averaged over a Hubble time. This is plotted in the upper two panels of Figure~\ref{fig1} as a function of redshift for $\epsilon_{\rm duty,tot}=0.1$ and 0.15 (solid lines), compared with observations of specific star formation rate at stellar masses of $M_\star=10^9$M$_\odot$ and $M_\star=5\times10^9$M$_\odot$ respectively. The dotted lines also show the curves corresponding to a duty-cycle of unity, illustrating the level at which the values of specific star formation rate deviate from expectations of continuous star formation. We note the weak dependence of the inferred duty-cycle over the range of stellar mass and redshift probed. This low duty cycle has a range of important implications for the properties of the high redshift galaxy population, and explains several puzzling properties of the observed relation between stellar mass and star formation rate. For the remainder of this paper we explore these explanations in the context of  the merger driven model of \citet[][]{Wyithe2013}. However we stress that the result of low duty-cycle from equation~(\ref{eq:simple}) is very general and not dependent on the details of our particular star formation model.

\section{Model}

\label{model}

A complication that arises when modelling the luminosity function is
that models predict a star formation rate, which must then be
converted to an observed luminosity assuming a dust extinction and an initial mass function (IMF) for
the stars. A better comparison between theory and observations is therefore to estimate the
star formation rate density observationally, where the correction is
made from luminosity to star formation rate using the observed
continuum properties of the galaxies under study. Following the approach of \citet[][]{Wyithe2013} we therefore focus on modelling the SFRD function \citep[][]{Smit2012} rather than the luminosity function of high redshift galaxies.

In this section we briefly summarise the model for star formation in high redshift galaxies presented in \citet[][]{Wyithe2013}. The reader is referred to that paper for further details of this model. The star formation rate in a galaxy halo of mass $M$ that turns a
fraction $f_\star$ of its disk mass $m_dM$ into stars over a time
$t_{\rm SF}$ is
\begin{equation}
\label{SFR}
SFR = 0.15 \mbox{M}_\odot \mbox{yr}^{-1}\left(\frac{m_{\rm
d}}{0.17}\right)\left(\frac{f_\star}{0.1}\right)\left(\frac{M}{10^8\mbox{M}_\odot}\right)\left(\frac{t_{\rm
SF}}{10^7\mbox{yr}}\right)^{-1} .
\end{equation}
The model assumes that major mergers trigger bursts of star formation. The star formation rate density
function (i.e. galaxies per Mpc$^{3}$ per unit of $SFR$) can be
estimated as
\begin{eqnarray}
\label{SFRD}
\nonumber
&&\hspace{-7mm}\Phi(SFR) = \\
&&\epsilon_{\rm duty}\left(\Delta M\left.t_{\rm H}\frac{dN^2_{\rm merge}}{dtd\Delta M}\right|_{M_1,\Delta M}\frac{dn}{dM}\right)\left(\frac{dSFR}{dM}\right)^{-1},
\end{eqnarray}
where $\epsilon_{\rm duty}$ is the fraction of the Hubble time
($t_{\rm H}$) over which each burst lasts, and $dn/dM$ is mass
function of dark matter halos \citep{Press1974,Sheth}. The rate of
major mergers ($dN_{\rm merge}/dt$) is calculated as the number of
halos per logarithm of mass $\Delta M$ per unit time that
merge with a halo of mass $M_1$ to form a halo of mass $M$
\citep[][]{Lacey1993}.  We assign a 2:1 mass ratio to major mergers
(i.e. $M_1=\frac{2}{3}M$ and $\Delta M=M/3$).  

The most massive stars fade
away on a timescale of $t_{\rm s}\sim3\times10^6$ years
\citep[][]{Barkana2001}. If the starburst lifetime is $t_{\rm SF}$ the duty-cycle can be written as
\begin{equation} 
\epsilon_{\rm duty} = \frac{t_{\rm s} + t_{\rm SF}}{t_{\rm H}}.
\end{equation}
For comparison with observations we define 
\begin{equation}
\Psi(SFR) = \ln{10}\times SFR \times \Phi,
\end{equation}
which has units of Mpc$^{-3}$ per dex.

We expect that SNe feedback will alter the fraction of gas in a galaxy
that is turned into stars~\citep[e.g.][]{Dekel2003}. To determine the
mass and redshift dependence of $f_{\star}$ in the presence of SNe we
suppose that stars form with an efficiency $f_{\star}$ out of the gas
that collapses and cools within a dark matter halo and that a fraction
$F_{\rm SN}$ of each supernova energy output, $E_{\rm SN}$, heats the
galactic gas mechanically (allowing for some losses due to
cooling). The mechanical feedback will halt the star formation once
the cumulative energy returned to the gas by supernovae equals the
total thermal energy of gas at the virial velocity of the halo
\citep[e.g.][]{Wyithe2003a}. Hence, the limiting stellar mass is set
by the condition
\begin{equation}
\label{feedback}
\frac{M_{\star}}{w_{\rm SN}}E_{\rm SN}F_{\rm SN}f_{\rm t}f_{\rm d}=E_{\rm b}=\frac{1}{2}m_{\rm d}Mv_{\rm vir}^2.
\end{equation}
In this relation $E_{\rm b}$ is the binding energy in the halo,
$w_{\rm SN}$ is the mass in stars per supernova explosion, and the
total stellar mass is $M_\star=m_{\rm d}\,M\,f_{\rm \star,tot}$ where
$f_{\rm \star,tot}=N_{\rm merge}f_{\star}$ is the total fraction of
the gas that is converted to stars during major mergers, and $N_{\rm
merge}$ is the number of major mergers per Hubble time. The parameters
$f_{\rm t}$ and $f_{\rm d}$ denote the fraction of the SNe energy that
contributes because of the finite timescale of the SNe feedback 
or the disk scale height being smaller than the SNe bubble. These
terms are described in more detail below.

The ratio between the total mass in stars and dark matter is observed
to increase with halo mass as $(M_\star / M)\propto M^{0.5}$ for
$M_\star \la 3 \times 10^{10}$M$_\odot$, but is constant for larger
stellar masses \citep[][]{Kauffmann2003}. Thus, the star formation
efficiency within dwarf galaxies decreases towards low masses.  For
comparison with equation~(\ref{feedback}), a \citet[][]{Scalo1998}
mass function of stars has $w_{\rm SN} \sim 126$ M$_\odot$ per
supernova and $E_{\rm SN} =10^{51}$ ergs, and so we find that
$M_\star=3\times10^{10}$ M$_\odot$ and $v_{\rm c} \sim 175$ km/s
\citep[the typical value observed locally; see e.g.][]{Bell2001}
implies $f_{\rm \star,tot}\sim 0.1$ for a value of $F_{\rm SN} \sim
0.5$. Smaller galaxies have smaller values of
$f_{\star}$. Equation~(\ref{feedback}) indicates that
\begin{eqnarray}
\label{fstar}
\nonumber
&&\hspace{-7mm}f_{\star}=\\
&&\hspace{-7mm}\min\left[f_{\rm \star,max},\frac{0.008}{N_{\rm merge}}\left(\frac{M}{10^{10}\mbox{M}_\odot}\right)^{\frac{2}{3}}\left(\frac{1+z}{10}\right)\left(f_{\rm t}f_{\rm d}F_{\rm SN}\right)^{-1}\right].
\end{eqnarray}
We utilise equation~(\ref{fstar}) with equation~(\ref{SFRD}) as a
function of the parameters $t_{\rm SF}$ and $f_{\rm \star,max}$.

\subsection{Disk structure}

The effect of SNe feedback is dependent on the conditions of the interstellar medium (ISM)
gas.  We assume that the cold gas (out of which stars form) occupies a
self-gravitating exponential disk where $R_{\rm d}$ is the scale radius, 
$m_{\rm d}$ is the mass fraction of the disk relative to the halo
and $\lambda\sim 0.05$ is the spin parameter of the halo \citep[][]{Mo1998}.
The scale height of the disk at radius $r$ is
\begin{equation}
H=\frac{c_{\rm s}^2}{\pi G \Sigma(r)},
\end{equation}
where $c_{\rm s}$ is the sound speed in the gas, which we assume to have a temperature of $10^4$K, and  $\Sigma(r)$ is the surface density. 
We adopt the density in the mid plane at the scale radius, within
which half the gas is contained, as representative of the density of
the ISM.

\begin{figure*}
\begin{center}
\vspace{-10mm}
\includegraphics[width=15cm]{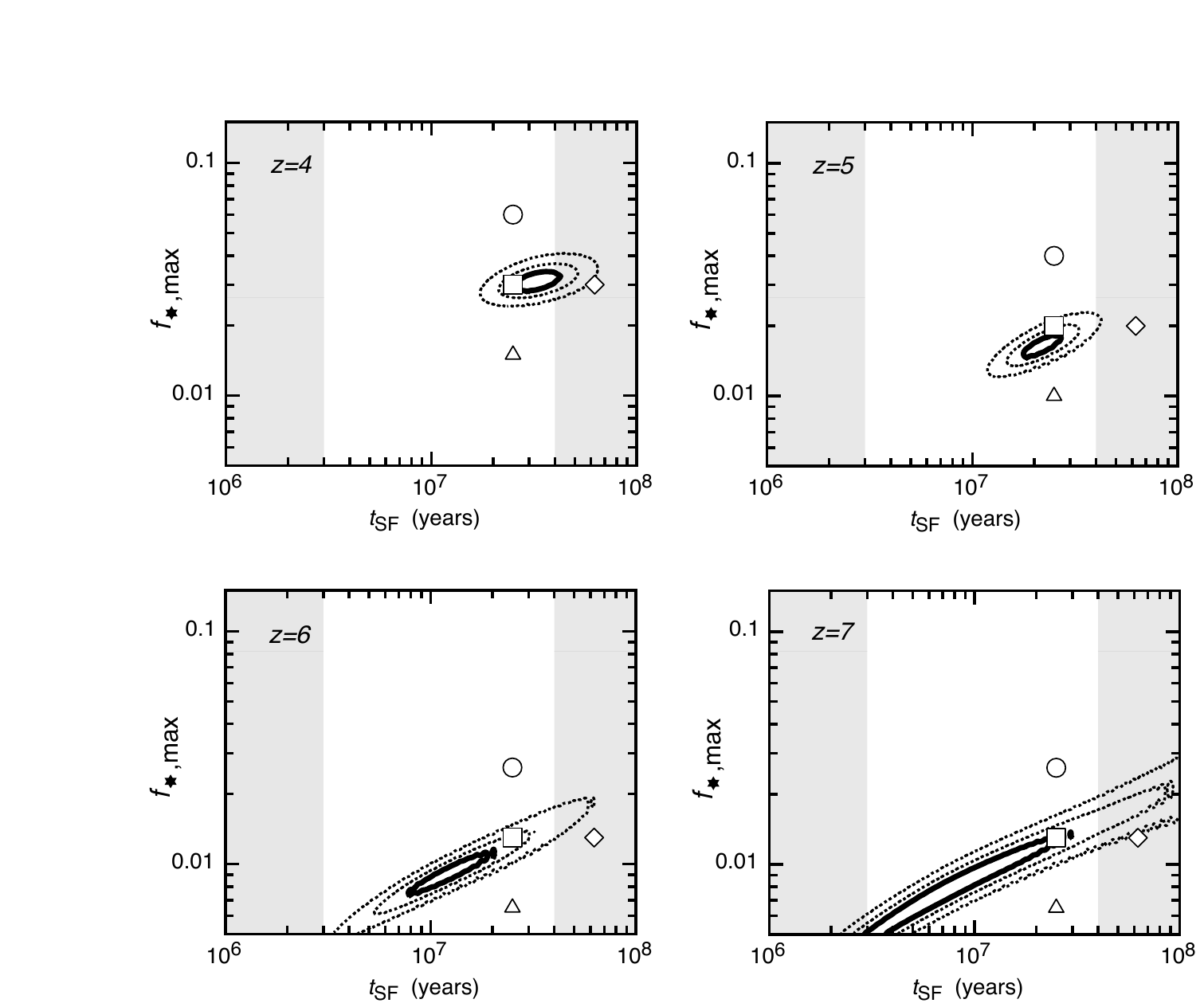}
\caption{\label{fig2}Constraints on the model parameters $f_{\rm
\star,max}$ and $t_{\rm SF}$ at four different redshifts (constraints
are independent at each redshift).  In each case, three contours are
shown corresponding to differences in $\chi^2$ relative to the
best-fitting model of $\Delta\chi^2 = \chi^2 - \chi^2_{\rm min} = 1$,
2.71 and 6.63. Projections of these contours on to the axes provide
the 68.3, 90 and 99 per cent confidence intervals on individual
parameter values. The vertical grey regions represent time-scales
longer/shorter than the lifetime of the highest/lowest mass SNe
progenitor ($3\times10^6$yr$/4\times10^7$yr). }
\end{center}
\end{figure*}

\subsection{Supernova evacuation of the ISM}

\citet[][]{Clarke2002} presented a simple analytic model for the
effect of supernovae on the interstellar medium which we apply to high
redshift galaxies. In this model, clusters of $N_{\rm e}$ SNe produce
super-bubbles in the ISM with a radius $R_{\rm e}$ at which the
super-bubble comes into pressure balance with the ISM. This radius can
be found by approximating $R_{\rm e}$ as the radius within which the
thermal energy of the ISM equals the mechanical energy of the SNe
cluster. The timescale associated with the evacuation of a super bubble in the ISM
 by a SNe cluster is $t_{\rm e}=4\times10^7$
years, corresponding to the lifetime of the lowest mass SNe progenitor.
 The evacuation radius for a cluster of $N_{\rm e}$ SNe, each
with energy output $E_{\rm SN}$ within an ISM of sound speed $c_{\rm
s}$ is
\begin{eqnarray}
\label{re}
\nonumber
R_{e} &=&0.08 \,\mbox{kpc} \left(\frac{N_{\rm e}}{10}\right)^{\frac{1}{3}} \left(\frac{E_{\rm SN}}{10^{51}\mbox{erg}}\right)^{\frac{1}{3}}  \left(\frac{\lambda}{0.05}\right)^{\frac{4}{3}} \left(\frac{m_{\rm d}}{0.17}\right)^{-\frac{2}{3}} \\
&&\hspace{5mm}\times\left(\frac{M}{10^8\mbox{M}_\odot}\right)^{-\frac{2}{9}} \left(\frac{1+z}{10}\right)^{-\frac{4}{3}}.
\end{eqnarray}

\begin{figure*}
\begin{center}
\vspace{3mm}
\includegraphics[width=17.5cm]{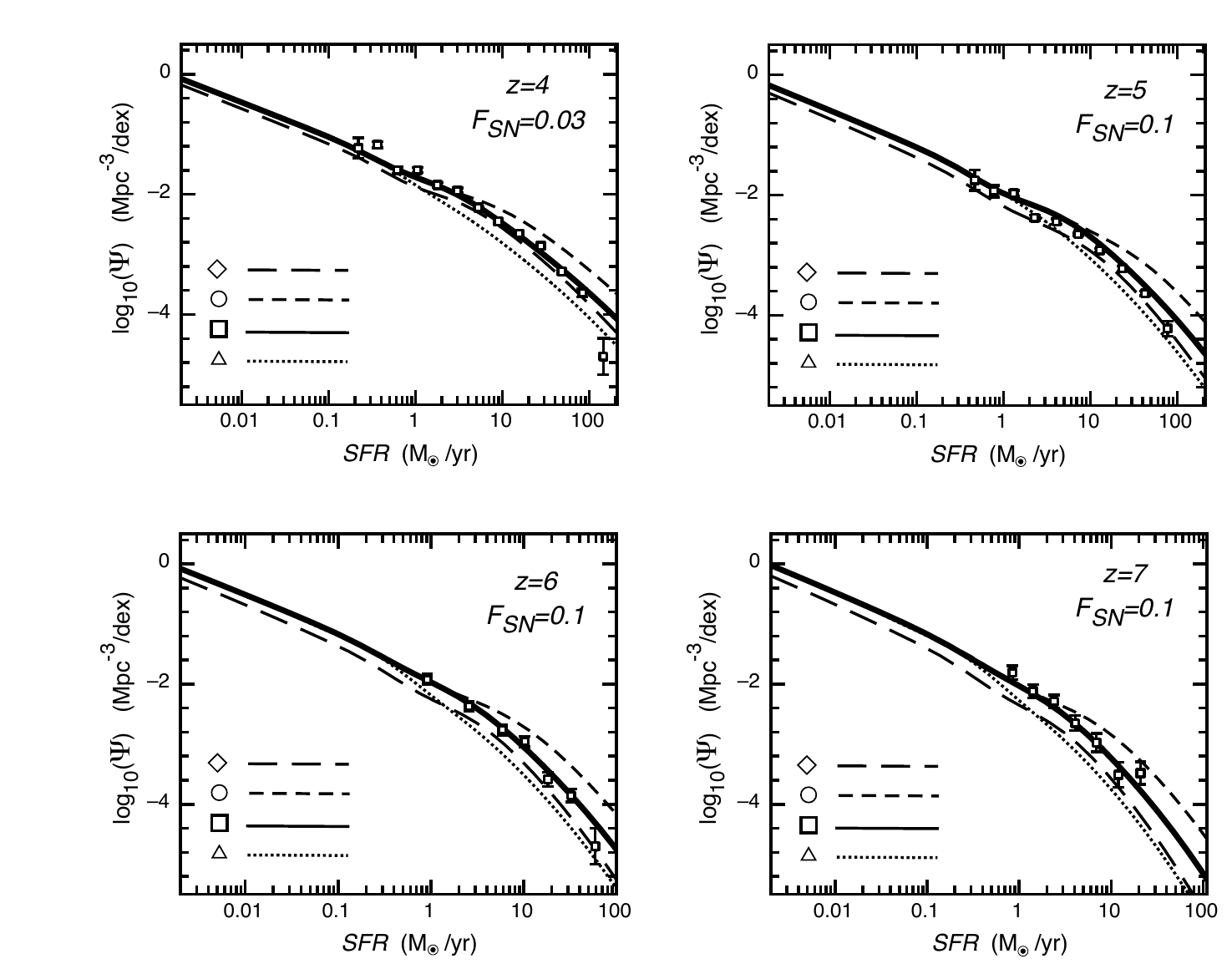}
\caption{\label{fig3}Comparison between the observed and modelled SFRD
function (plotted as $\Psi = \ln{10}\times SFR\times\Phi$). The four panels
show results for different redshift values. The observed SFRD functions from \citet{Smit2012} are shown as small open squares.  In each panel, the four
curves correspond to a different choice of model parameters $t_{\rm
SF}$ and $f_{\rm \star,max}$, labelled by the symbols in
Figure~\ref{fig2}. The thick solid lines represent values close to the
best fit. }
\end{center}
\end{figure*}

In the limit where SNe evacuated regions are smaller than the scale
height of the disk, and the starburst lifetime $t_{\rm SF}$ is much
larger than the gas evacuation timescale $t_{\rm e}$, the fraction
$F_{\rm SN}$ of the SNe energy may be used in feedback suppressing
subsequent star formation. However, if the SNe evacuated regions break
out of the disk, or $t_{\rm SF}<t_{\rm e}$, not all of the energy will
be available for feedback. Based on the ISM porosity model of
\citet[][]{Clarke2002}, a fraction $f_{\rm d}=2H/R_{\rm e}$ of the SNe
energy goes to increasing the ISM porosity for disks where $R_{\rm
e}>H$. In this case we find
\begin{eqnarray}
\nonumber
f_{\rm d}&=&0.85 \left(\frac{N_{\rm e}}{10}\right)^{-\frac{1}{3}} \left(\frac{E_{\rm SN}}{10^{51}\mbox{erg}}\right)^{-\frac{1}{3}}  \left(\frac{\lambda}{0.05}\right)^{\frac{2}{3}} \left(\frac{m_{\rm d}}{0.17}\right)^{-\frac{1}{3}} \\
&\times&\hspace{5mm}\left(\frac{M}{10^8\mbox{M}_\odot}\right)^{-\frac{1}{9}} \left(\frac{1+z}{10}\right)^{-\frac{2}{3}}\left(\frac{c_{\rm s}}{10\mbox{km/s}}\right)^2,
\end{eqnarray}
as long as $f_{\rm d}<1$ and $f_{\rm d}=1$ otherwise. Similarly, in
cases where $t_{\rm SF}<t_{\rm e}\sim 4\times10^7$ yrs, only
\begin{equation}
f_{\rm t}\equiv(t_{\rm SF}/t_{\rm e})^2
\end{equation}
of the overall SNe energy output is generated by the time the
starburst concludes. The quadratic dependence on time arises because
the number of bubbles produced grows in proportion to time, while the
maximum size of a bubble at time $t<t_{\rm e}$ is also proportional to
time \citep[][]{Oey1997}. In cases where $t_{\rm SF}>t_{\rm e}$ we
have $f_{\rm t}=1$.

\section{Results}
\label{results}

\begin{figure*}
\begin{center}
\vspace{3mm} \includegraphics[width=15cm]{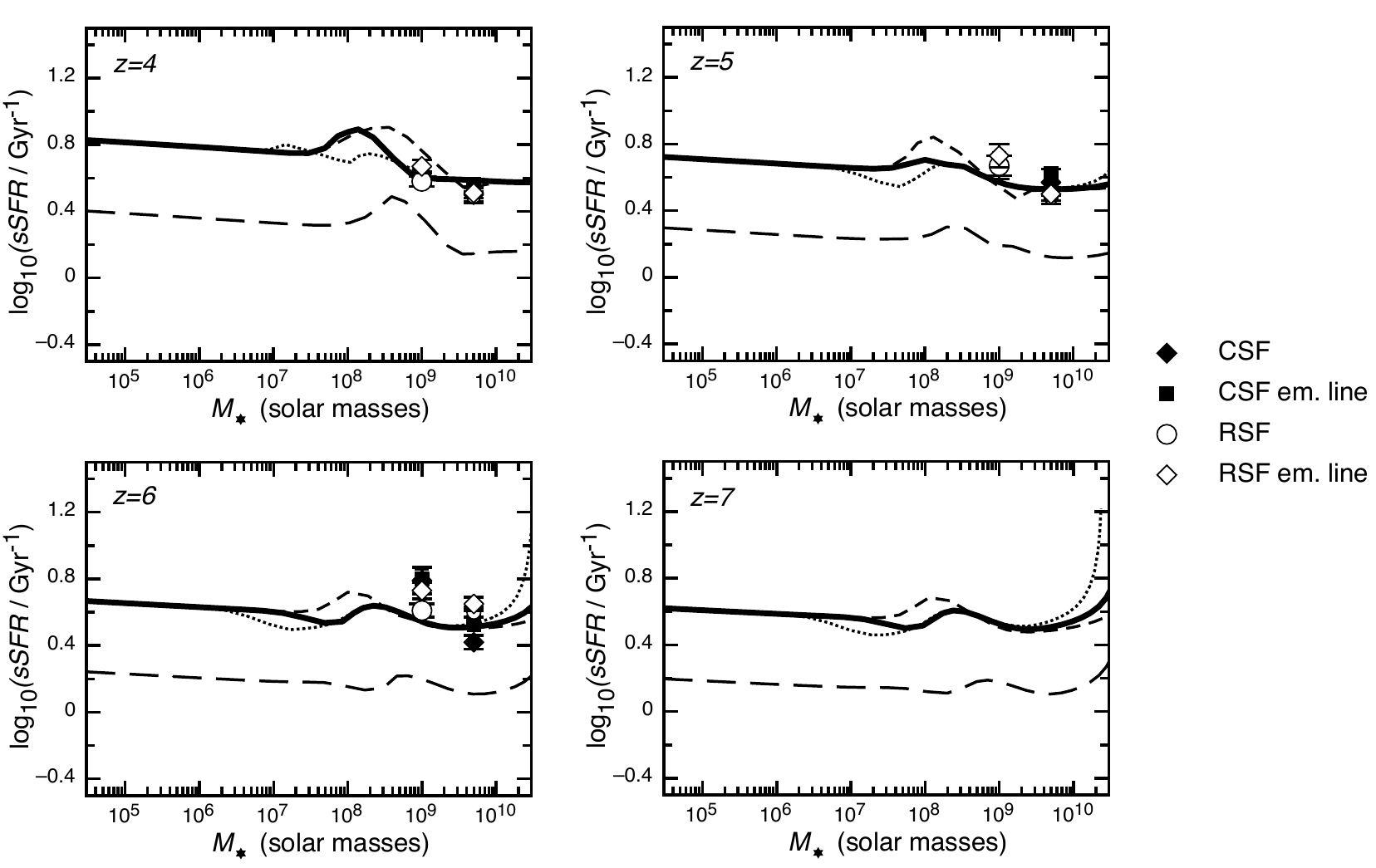}
\caption{\label{fig4} The specific star-formation rate for
the models in Figure~\ref{fig3} as a function of stellar mass (as predicted by the
star-burst only model).  The data points are from  \citet[][]{Gonzalez2012}, and the labels represent stellar masses based on the assumptions of a constant (CSF) and rising (RSF) star-formation history without and with emission lines included (em. line) as described in that paper.}
\end{center}
\end{figure*}

\begin{figure*}
\begin{center}
\vspace{3mm} \includegraphics[width=15cm]{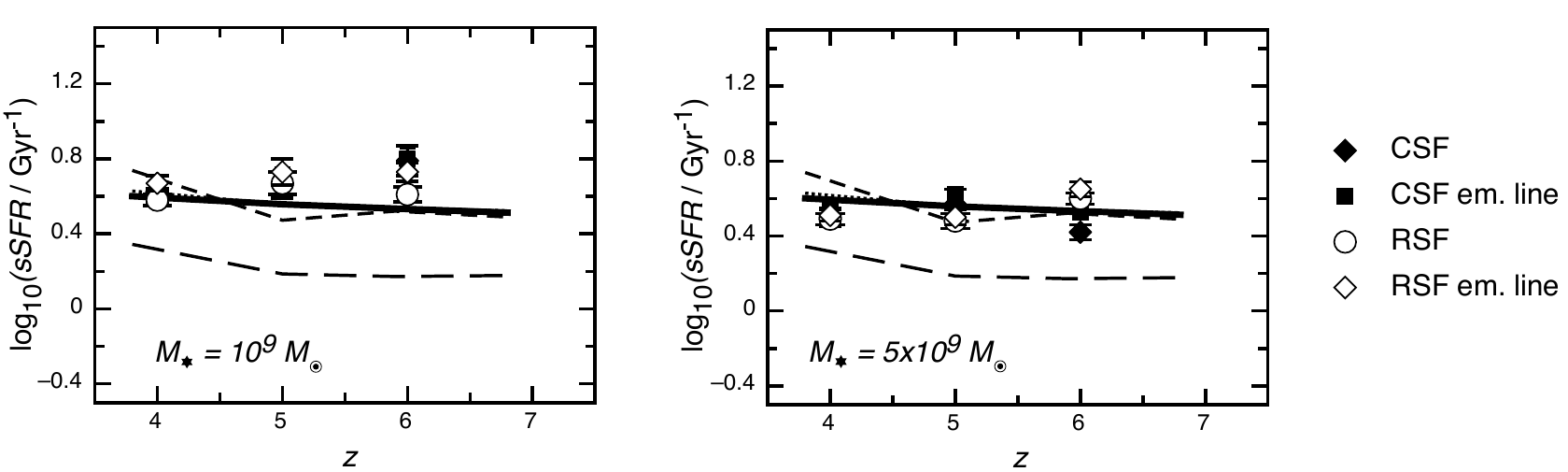}
\caption{\label{fig5} The specific star-formation rate for the models in Figure~\ref{fig3} as a
function of redshift at stellar masses of $10^9$M$_\odot$ and $5\times10^9$M$_\odot$. The data points are from  \citet[][]{Gonzalez2012}, and the labels represent stellar masses based on the assumptions of a constant (CSF) and rising (RSF) star-formation history without and with emission lines included (em. line) as described in that paper.}
\end{center}
\end{figure*}

\citet[][]{Wyithe2013} used the merger driven model described above to argue that the duty cycle for star formation in high redshift galaxies was a few percent. However, while much lower than unity the model independent estimate based on the specific star formation rate shown in Figure~\ref{fig1} is significantly larger, with a value of $\sim0.1$. To understand the origin of this difference we note that there is a degeneracy in the model between the duty-cycle $\epsilon_{\rm duty}$ and the parameter $F_{\rm SN}$ which governs the fraction of SNe energy that is harnessed for feedback. \citet[][]{Wyithe2013} arbitrarily chose a value for this parameter since it is unconstrained by just the SFR density function. However, for the analysis in this paper the inclusion of the constraint on specific star formation rate allows us to constrain the value of $F_{\rm SN}$ in addition to the values of $f_{\rm \star,max}$ and $t_{\rm SF}$.

\subsection{Comparison with observations}

We fit our model to the recent determination of the SFRD function from
\citet[][]{Smit2012}, and the specific star formation rate from \citet[][]{Gonzalez2012} in order to constrain the three free
parameters of our star formation model $t_{\rm SF}$ and
$f_{\star}$. We fit separately for four different redshifts $z\sim4$,
5, 6 and 7. Specifically we use the model to calculate SFRD functions and sSFR for
combinations of these parameters and calculate the $\chi^2$ of the
model as
\begin{eqnarray} 
\nonumber
\chi^2(f_{\rm \star,max},t_{\rm SF})&=&\\
\nonumber
&&\hspace{-30mm}\sum_{i=0}^{N_{\rm obs}}\left(\frac{\log{\Psi(SFR_i,f_{\rm \star,max},t_{\rm SF},z)}-\log{\Psi_{\rm obs}(SFR_i,z)}}{\sigma_{SFR}(SFR_i,z)}\right)^2\\
&&\hspace{-30mm}+\left(\frac{\log{sSFR(f_{\rm \star,max},t_{\rm SF},z)}-\log{sSFR_{\rm obs}(z)}}{\sigma_{sSFR}(z)}\right)^2
\end{eqnarray}
Here $\Psi_{\rm obs}(SFR_i,f_{\rm \star,max},t_{\rm SF},z)$ is the
observed star formation rate density measured at redshift $z$, with
uncertainty in dex of $\sigma_{SFR}(SFR_i)$. In calculating
likelihoods at $z\sim4$ and $z\sim5$ we increased the quoted error
bars by factors of 3 and 2 respectively in order to obtain a reduced
$\chi^2$ of order unity. The value of sSFR is evaluated at a stellar mass of $M_{\star}=10^9$M$_\odot$, with uncertainties corresponding to the range of estimates for the two assumed star formation histories and the two cases of with/without emission lines considered in \citet[][]{Gonzalez2012}. The SFRD function and sSFR are sensitive to the value
of $F_{\rm SN}$, 
and we therefore integrate the likelihood over a range of values
uniformly distributed between $-2.5<\log_{10}{F_{\rm SN}}<0$
\begin{equation}
\mathcal{L}(f_{\rm \star,max},t_{\rm SF})\propto\int_{-1}^0 d(\log_{\rm 10}{F_{\rm SN}}) e^{-\chi^2/2}.
\end{equation}

We note that the relation between $SFR$ and $M$ in
equation~(\ref{SFR}) is not perfect. As part of our comparison with
observations, and to account for scatter in this relationship, we
convolve the predicted SFRD function from equation~(\ref{SFRD}) with a
Gaussian of width $0.5$ dex in $SFR$.  An intrinsic scatter of 0.5 dex
is motivated by the scatter in stellar mass at constant $SFR$ found by
\citet[][]{Gonzalez2011}. However, in addition we find that the value
of 0.5 dex provides the best statistical fit to the observations. Our
qualitative results are not sensitive to the choice of this scatter.  

\begin{figure*}
\begin{center}
\vspace{3mm}
\includegraphics[width=15cm]{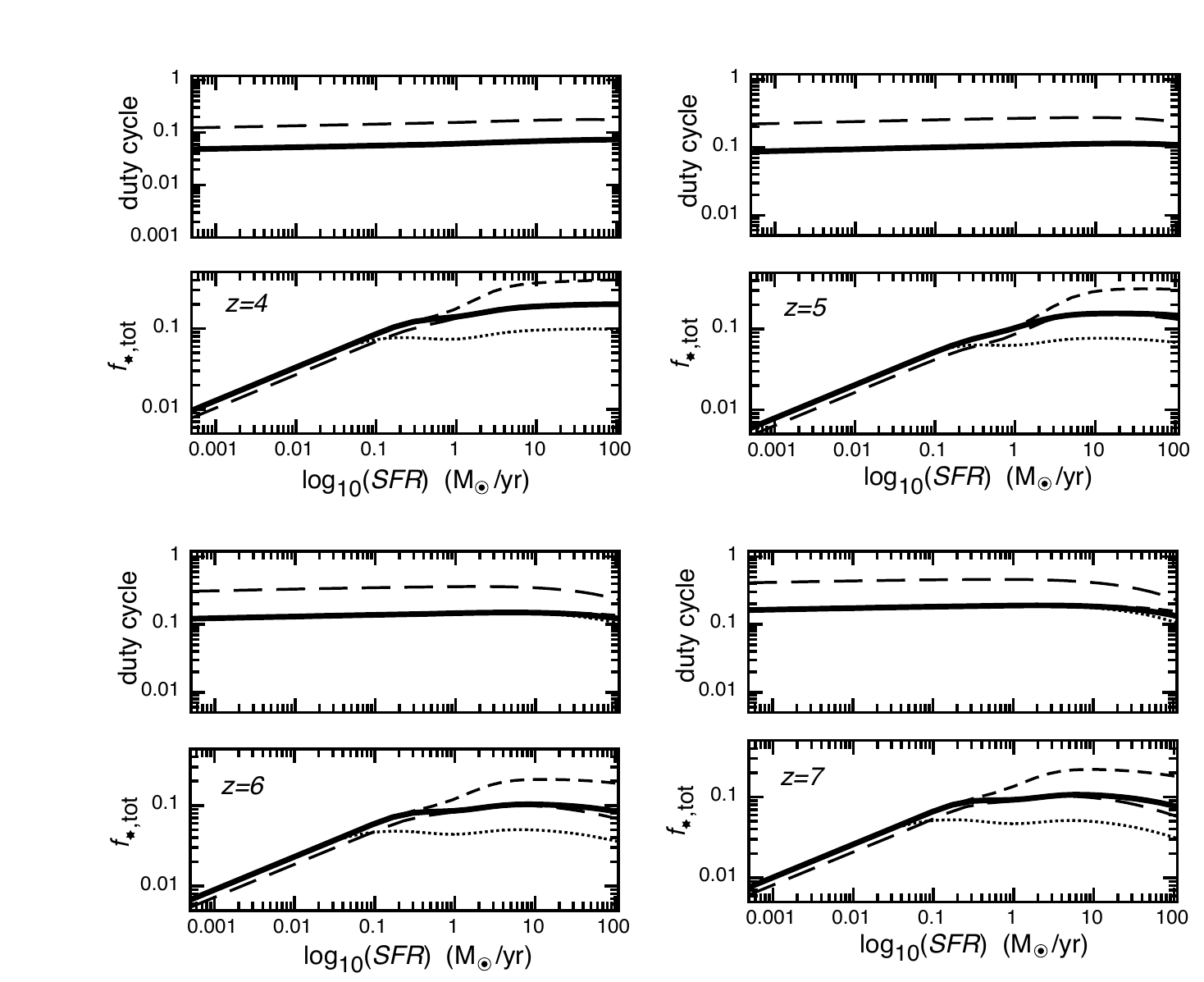}
\caption{\label{fig6}The values of total star formation efficiency
$f_{\rm \star,tot}$ (i.e. the sum of $f_{\rm \star}$ over all
mergers), and the overall duty-cycle (i.e. the fraction of a Hubble
time during which a galaxy is star bursting) as a function of $SFR$.
The four curves shown correspond to the SFRD functions shown in
Figure~\ref{fig3}, with model parameters $t_{\rm SF}$ and $f_{\rm
\star,max}$ designated by the symbols in Figure~\ref{fig2}. }
\end{center}
\end{figure*}

\subsection{Parameter constraints}

In Figure~\ref{fig2} we show constraints on models.  
Given these sSFR driven constraints on $F_{\rm SN}$,  we find that the shape of the SFRD function
requires starburst durations of a few tens of Myr
at $z\sim5$, $z\sim6$ and $z\sim7$, with a few percent of the gas
turned into stars per burst. For comparison the left and right hand
vertical grey regions represent times smaller than the lifetime of the
most massive stars ($t_{\rm s}\sim3\times10^6$ years), and times in
excess of the lifetime of the least massive stars that produce SNe
respectively. Our results therefore indicate that star formation in
high redshift galaxies is terminated on the same timescale as feedback
from SNe can be produced \citep[][]{Wyithe2013}.  

Figure~\ref{fig3} shows the comparison
between observed and modelled SFRD functions for four different redshifts $z\sim4$,
5, 6 and 7 with values of $F_{\rm SN}=0.03$ at $z=4$ and $F_{\rm SN}=0.1$ at $z=5,6$ and 7 (c.f. $F_{\rm SN} = 0.1$ and 0.3 in \citet[][]{Wyithe2013}).  The
burst lifetime is $t_{\rm SF}=2.5\times10^7$ years in each case. The four curves shown
correspond to model parameters $t_{\rm SF}$ and $f_{\rm \star,max}$
labeled by the symbols in Figure~\ref{fig2}. The thick solid lines show models close to the
best fit to the observational data\footnote{These curves are not plotted at the formal best fit because whereas the constraints were determined independently, we have chosen common values for parameters $f_{\rm \star,max}$ and
$t_{\rm SF}$ across several redshifts.}.  The other three values were chosen so as to illustrate the dependence of the predicted SFRD function on the different parameters. 

\subsection{Specific star formation rate}

Figure~\ref{fig4} shows the specific star formation rate as a function of mass for the models listed in Figure~\ref{fig3} at $z=4$, 5, 6 and 7, illustrating the success of the model in reproducing the observed specific star formation rate for the constrained parameters (particularly $F_{\rm SN}$). Beyond the narrow range of observed stellar mass values, the model predicts that the specific star-formation rate remains quite insensitive to stellar mass (or star formation rate). In this figure the wiggles result from the simple assumption of an abrupt change in efficiency with halo mass in equation~(\ref{fstar}). Figure~\ref{fig5} shows the specific star formation rate as a function of redshift for these models, illustrating the model prediction that the specific star-formation rate does not evolve with redshift. This finding is in agreement with observations, but in contrast to results from many hydrodynamical models of galaxy formation, indicating that star formation activity does not directly follow the gas accretion rate.

\subsection{Star formation efficiency and average duty-cycle}

The parameters $f_{\rm \star,max}$ and $t_{\rm SF}$ refer to single
bursts, whereas our model includes multiple bursts occurring at the rate of
major mergers. We therefore calculate the total star formation
efficiency $f_{\rm \star,tot}=H^{-1}\,dN_{\rm merge}/dt\, f_\star$ (i.e. the sum of
$f_{\rm \star}$ over all mergers during a Hubble time), as well as the overall duty-cycle\footnote{We note that we calculate the total duty-cycle and star formation efficiency centred on a particular redshift, whereas observationally measurements are made over a redshift range. This approximation is justified owing to the slow evolution in merger rate.}
$\epsilon_{\rm duty,tot}= N_{\rm merge}t_{\rm SF}/t_{\rm H} = t_{\rm SF}\,dN_{\rm merge}/dt$.  These quantities are plotted in
Figure~\ref{fig6} based on our model with parameter choices
corresponding to the examples in Figure~\ref{fig3}. We find that $\sim5-10\%$ of
the gas forms stars in bright galaxies of $SFR\sim1-100$M$_\odot$ per year,
with lower fractions down to a percent in fainter galaxies. This trend is in qualitative agreement with the observations of \citet[][]{Lee2012} to explain the stellar mass vs UV luminosity relation. We find
duty-cycles of $\sim10-20$\%, with higher duty-cycles at higher
redshift reflecting the increased ratio between the lifetime of
massive stars and the age of the Universe. The duty cycle is also larger for systems of higher star formation rate. This trend is in agreement with the observational estimate of \citet[][]{Lee2009} based on a comparison of the luminosity and clustering of luminous $z\ga4$ galaxies. These authors find that star-formation is constrained to be bursty, and infer a duty-cycle at $z\sim4$ of 15\%-60\% (at 1-$\sigma$).

\subsection{The stellar mass function}

\begin{figure*}
\begin{center}
\vspace{3mm} \includegraphics[width=15cm]{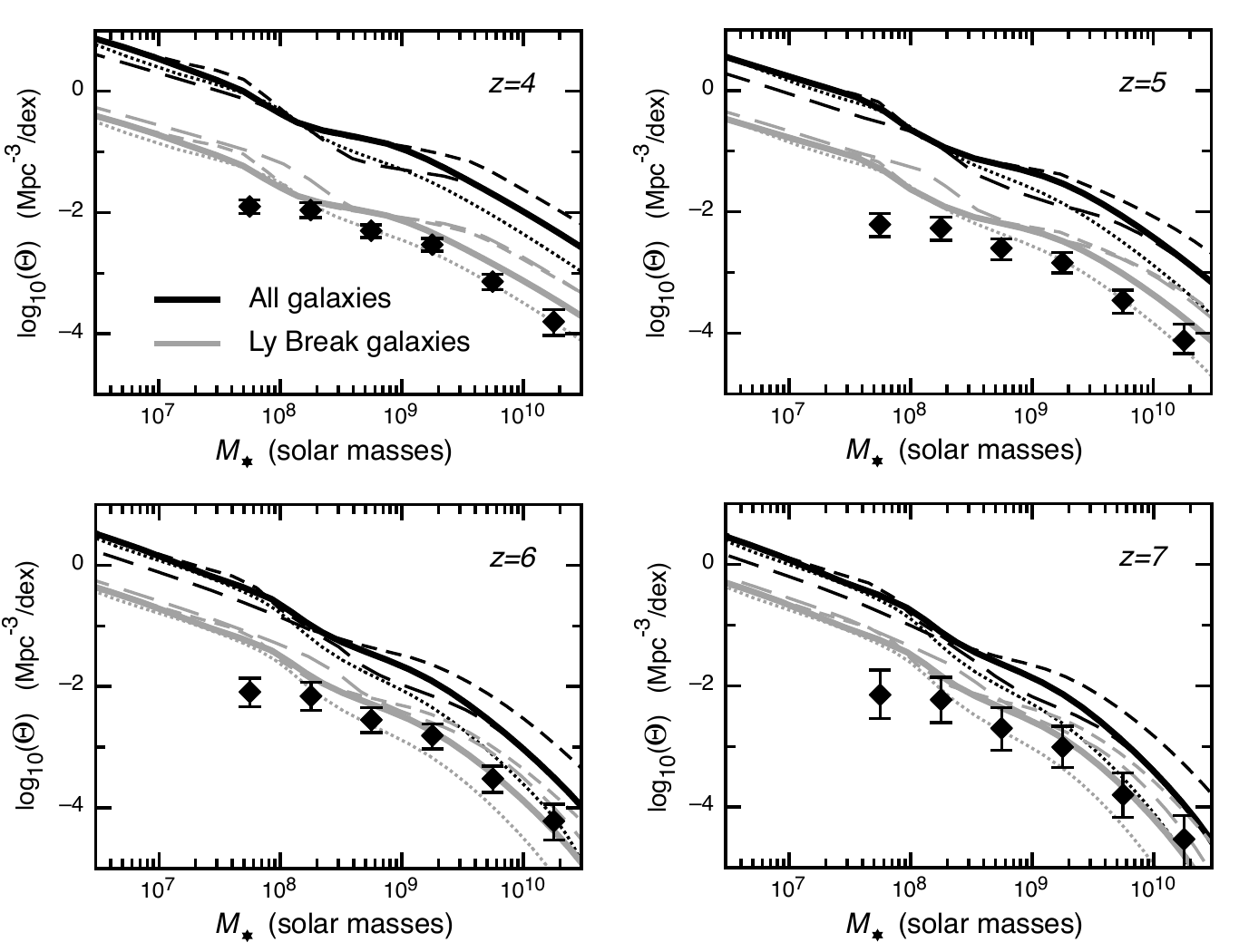}
\caption{\label{fig7} The stellar mass function of a Ly-break selected sample. This is plotted as the grey curves in the attached figure. These curves are for the same models as Figures~\ref{fig3}-\ref{fig8}. The data points are from  \citet[][]{Gonzalez2011}. We also calculate the mass function that would be seen if the sample were selected in stellar mass rather than SFR. The black curves show these models, which are a factor of 5 to 10 higher.}
\end{center}
\end{figure*}

\begin{figure*}
\begin{center}
\vspace{3mm} \includegraphics[width=15cm]{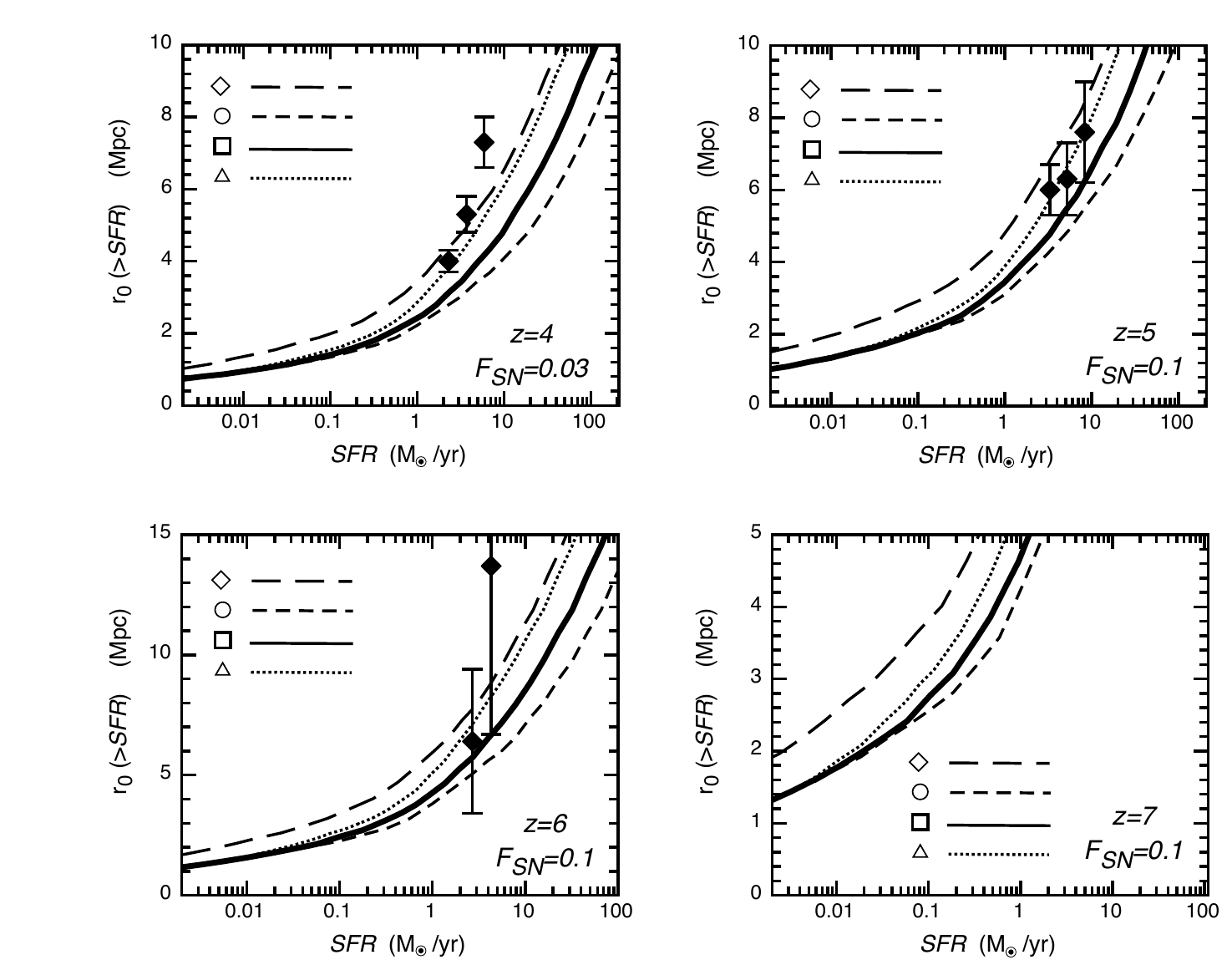}
\caption{\label{fig8} The correlation length in
samples above a limiting SFR. The data points are from \citet[][]{Lee2009} and
\citet[][]{Overzier2006}. The correlation length was calculated for each sample
above the limiting star-formation rate by averaging over
correlation functions weighted by the SFR density function. Clearly, our model predictions are in good agreement with current observational clustering measurements at $z\sim5-6$, providing an independent verification of the viability of the model. At $z\sim4$, our models underestimate the clustering strength somewhat, which may indicate that a smooth contribution to the star formation is required at lower redshifts in a population of more massive, biased halos.}
\end{center}
\end{figure*}

The next observable that we consider is the stellar mass function. Observationally, the stellar mass function at $z\gtrsim4$ is currently derived for star-forming, UV-bright, Ly-break selected galaxies. This can be estimated from our model as
\begin{equation}
\label{SMF}
\Theta(M_\star) = \Phi(SFR) \times \left(\frac{dSFR}{dM_\star}\right) =  \Phi(SFR) \times sSFR\\.
\end{equation}
The resulting stellar mass function is plotted as the grey curves in Figure~\ref{fig7} for the models shown in Figure~\ref{fig3}-\ref{fig8}. The data points are from  \citet[][]{Gonzalez2011}. Since our model produces both the correct specific star formation rate and the correct SFRD function, it is no surprise that the agreement is good. This agreement is in contrast to many hydrodynamical models of galaxy formation, in which the constant accretion leads to high duty cycles, and hence a SFR to halo mass ratio which is too low. This low mass-to-light ratio in turn results in a stellar mass function that is too steep. 

The low duty-cycle of our bursty model predicts the existence of many galaxies of large stellar mass that are not star forming, and which may therefore be missing from a Ly-break selected stellar mass function. Using our model we therefore also calculate the predicted mass function in the case where the sample was selected based on stellar mass rather than on SFR. This is 
\begin{eqnarray}
\label{SMF2}
\nonumber
\Theta_{\rm all}(M_\star) &=& \frac{1}{\epsilon_{\rm duty,tot}}\Phi(SFR) \times \left(\frac{dSFR}{dM_\star}\right) \\
\nonumber
&=&  \frac{1}{\epsilon_{\rm duty,tot}}\Phi(SFR) \times sSFR \\
&=&\frac{1}{\epsilon_{\rm duty,tot}}\times\Theta(M_\star).
\end{eqnarray}
In Figure~\ref{fig7} we show the resulting predicted stellar mass functions for the models shown in Figure~\ref{fig3}-\ref{fig8} (black curves). Owing to the low duty-cycle, these curves are a factor of $\sim5-10$ higher than the Ly-break selected case, indicating that high redshift surveys may currently be missing most of the stellar mass produced at early times.

\subsection{Clustering of star forming galaxies}

Finally, to check whether the relationship between halo mass and SFR is correctly reproduced in our model we calculate the correlation length in samples above a limiting SFR for the four models in this paper. The results are plotted in Figure~\ref{fig8}.
The correlation length is calculated for a sample above the limiting star-formation rate by averaging over correlation functions weighted by the SFR density function. For comparison we also include clustering measurements from Lee et al. (2009) and Overzier et al. (2006). To convert from an apparent magnitude limit to an intrinsic star formation rate, we assumed a flat SED with $\beta=-2$ for computation of a K-correction, and a conversion from UV luminosity to SFR using Kennicutt (1998). We find that the clustering length increases rapidly towards high SFRs, in agreement with observations. Our model yields a clustering length in the best fit model which is consistent with clustering measurements at $z\sim5-6$, but underestimates observations at $z\sim4$. This underestimate may indicate that a smooth contribution to the star formation is required at lower redshifts which would require a population of more massive, biased halos.

\section{The Observability of a Passive High-Redshift Population}
\label{lowlum}

As outlined above, due to the low effective duty cycle, our model predicts a significant population of galaxies that is not star-forming at a given point in time and might therefore be missed by current high-redshift surveys, which are only sensitive to the rest-frame UV. An accurate estimate for the extent of this missing population should be derived through detailed spectral energy distribution modelling of the whole galaxy population, which is beyond the scope of this paper. However, we briefly outline the likely impact of a low duty cycle and bursty star formation history on the observability of the full galaxy population at high redshift.

There are two ways for a non-starforming galaxy at high-redshift to be missed in a Lyman-break selected sample. Firstly, because the UV luminosity could dim below the detection limit of a survey, and secondly the UV continuum colour could evolve too far to the red where the effective survey volume of a typical LBG selection drops significantly for sources with UV continuum slopes $\beta\gtrsim-1$. In practice however, the drop in the UV luminosity is likely the main cause for a galaxy not being selected. Based on simple star-formation histories with short star-formation times ($t_\mathrm{SF}<100$ Myr) and using \citet{Bruzual2003} stellar population models, one finds that a galaxy dims by an order of magnitude (i.e. 2.5 mag) in the rest-frame UV light after $\lsim100$ Myr of a passive phase. Given the star formation lifetime of $t_{\rm SF}\sim10^7$ yr and duty-cycle of $\sim10\%$, this value of $\sim100$ Myr is comparable to, but shorter than the average time between bursts in our model at $z>4$, leading to sources being missed from current surveys. Furthermore, after 100 Myr, a galaxy's rest-frame UV continuum slope would have reddened by $\Delta\beta>1$, which would further diminish its chance to be selected as a robust high-redshift source.

Our model therefore predicts a significant population of UV faint galaxies with red UV continuum slopes of $\beta\gtrsim-1$. Such sources could in principle be searched for based on $Spitzer$/IRAC imaging which samples rest-frame optical wavelengths of $z\sim4-8$ sources. In the rest-frame optical, a galaxy would only dim by $\sim1.5$ mag after 100 Myr without star formation. Such galaxies would be detectable in the deepest current IRAC images over the Hubble Ultra-Deep Field \citep[reaching down to $\sim27$ mag$_\mathrm{AB}$; see][]{Oesch2013} out to $z\sim6$ if their stellar masses are a few times 10$^9 M_\odot$. However the selection of such galaxies is complicated by possible contamination from very dusty low-redshift sources \citep[see e.g.][]{Caputi2012,Wiklind2008}, as well as by confusion due to the broad IRAC point-spread function. The advent of JWST will, therefore, greatly facilitate the identification of passive high-redshift sources due to its higher resolution and much better sensitivity in several filters sampling the observed $>2~\mu$m regime, and will for the first time enable a full census of high-redshift galaxies.

\section{Discussion}
\label{discussion}

In this paper we have described a bursty model for SNe regulated high redshift star formation. Our model successfully reproduces the star-formation selected stellar mass function because it predicts both the star formation rate density function and the correct specific star formation rate of star forming galaxies. However, if we are considering the stellar mass function of the whole galaxy population then there is stellar mass missing from the observed census. Indeed, based on our model, we argue that surveys currently find only $\sim10-20\%$ of the total stellar mass density at $z\ga4$.We argue that the low duty-cycle of star formation in this model produces possible solutions to two observed puzzles in high redshift galaxy formation. 

The first puzzle  relates to the observation that the specific star-formation rate does not evolve significantly with redshift or mass at $z\geq4$, in contrast to theoretical expectation. Our bursty model naturally produces this behaviour. We point out that the value of the specific star formation rate, and its observed evolution at high redshift directly constrain the duty-cycle (averaged over a Hubble time) of high redshift star-formation to be approximately 10\%, independent of a specific model for star-formation. 

The second puzzle lies in the relation between the observed growth of stellar mass and the observed instantaneous star formation rate. The stellar mass density that is directly observed in samples at high redshift is the stellar mass density in the population of star forming galaxies \citep[][]{Gonzalez2011}. This quantity may be different from the total stellar mass density in the Universe. There seems to be disagreement between the relationship of star formation rate to stellar mass observed at $z\sim6$ \citep[][]{Bouwens2011} and at $z\sim2-4$ \citep[][]{Wilkins2008}. Specifically, at $z\sim2-4$,  \citet[][]{Wilkins2008} find that there is not enough growth in stellar mass to account for all of the star-formation observed.  \citet[][]{Wilkins2008} calculate the stellar mass density ($\rho_{\rm \star, obs}$) using fits to the stellar mass function, and take the derivative across a redshift interval $\Delta z\sim0.5$ to find an inferred star formation rate $\dot{\rho}_{\rm \star,inf}$ in units of mass per time per Mpc$^3$. Comparing with the observed star formation rate $\dot{\rho}_{\rm \star,obs}$ at $z\sim2-4$, \citet[][]{Wilkins2008} found that $\dot{\rho}_{\rm \star,obs}  > \dot{\rho}_{\rm \star,inf}$  with a difference of $\sim0.6$ dex.  However with a duty-cycle smaller than unity, the stellar mass in the star forming galaxies was built up over a time shorter than the survey depth, meaning that $\rho_{\rm \star,inf}$ is an overestimate relative to the observed stellar mass. In this case only a fraction $\epsilon_{\rm duty}$ of galaxies with stellar mass $M_\star$ are observed in a particular survey, but all galaxies would have starbursts during a time corresponding to the survey depth (note this does not imply that the instantaneous SFRD is underestimated). Thus, if a Ly-break is needed to identify the galaxies in which stellar mass is observed, much of the stellar mass at a particular time would be missed by the survey since it is contained in non-star forming galaxies. The difference is a factor of inverse the duty-cycle, explaining the disagreement between observed and inferred quantities found by \citet[][]{Wilkins2008}. We note that if the galaxy sample were selected on stellar mass rather than on UV luminosity, the estimates of star formation rate density based on instantaneous SFRD and the derivative of stellar mass density would agree. 

In a complementary analysis \citet[][]{Bouwens2011} have taken the stellar mass function at $z\sim6-8$ determined by \citet[][]{Gonzalez2011} and differentiated to get the star formation rate in a survey at $z>6$. However, in contrast to the results of  \citet[][]{Wilkins2008} at $z\sim2-4$, in this higher redshift case the resulting stellar mass is found to agree well with the stellar mass inferred directly. At first sight this is a failure for our model, which predicts that these estimates should differ by a factor of inverse duty-cycle as they do at lower redshift. The solution to this apparent contradiction lies in the fact that at $z\sim6-8$ the survey depth of $\Delta z\sim0.5$ corresponds to a time difference across the survey that is similar to the starburst lifetime (but shorter than the time between major mergers), in contrast to the case at $z\sim2-4$ where $\Delta z\sim0.5$ corresponds to a time that is longer than the star-burst lifetime. 

Thus, in difference to observations at $z\sim2-4$, at $z\ga6$ we expect that the observed galaxies had star formation episodes for a time that is similar to the survey depth, meaning that the stellar mass census within star forming galaxies does include all the stellar mass that was generated during the survey depth time interval. As a result, in the $z\ga6$ samples, integrating the observed star formation rate between the upper and lower redshifts of the survey gives a stellar mass that approximately equals the mass observed in those $z\sim6$ galaxies, in agreement with the comparison of \citet[][]{Bouwens2011} \citep[see also][]{Robertson2013}. However this equivalence is a coincidence, and does not correspond to a large duty cycle. Rather, galaxies that are not star forming, and therefore not seen in the survey did not form stars during the survey interval. Thus as in the $z\sim2-4$ case, there is additional stellar mass in quiescent galaxies that is not accounted for in the observed stellar mass function.

We note that this finding is in contrast to recent work at $z\la3$ that has used the UltraVISTA survey to construct a K-selected catalog covering masses $M_\star\ga10^{11}$M$_\odot$ \citep[][]{Muzzin2013,Ilbert2013}. These authors find stellar mass functions selected by stellar mass to agree reasonably well with UV selected samples, and that star-forming galaxies dominate the stellar mass density at low redshift. This would indicate that $z\la3$ UV selections are not necessarily missing too many galaxies at the very massive end of the population. At $z\ga4$ our model is not consistent with this behaviour, which is indicative of a larger duty-cycle or a component of continuous star-formation. Full SED modelling of a galaxy population based on merger trees will be required to understand all the observational selection effects on the observability of the predicted non-star forming galaxy population from our model in detail.

\section{Conclusion}

We have shown that a bursty model of high redshift star formation reproduces a range of observations of the high redshift galaxy population. In particular, we point out that the observed specific star formation rate requires a duty-cycle of $\sim10\%$, which follows directly from the fact that the observed star formation rate in galaxies integrated over a Hubble time exceeds the observed stellar mass by an order of magnitude.  We use this observational constraint to calibrate the efficiency of feedback in a model for the high redshift star formation rate which includes merger driven star formation regulated by SNe feedback. This model reproduces the star formation rate density function and the stellar mass function of galaxies at $4\la z\la7$. 

The finding of a $\sim10\%$ duty cycle implies that there are ten times the number of known galaxies at fixed stellar mass that have not yet been detected. Since current observations select galaxies by their UV luminosity, and hence only detect star-forming galaxies without too much extinction, our model predicts a large undetected population of UV-faint galaxies that accounts for most of the stellar mass density at $z=4-8$. Unfortunately, at $z>4$ there are currently no good constraints on non star-forming galaxies. These UV faint galaxies would be detectable through their rest-frame optical emission with the Spitzer Space Telescope or JWST. However, it is difficult to define selection criteria that select such sources without significant contamination from lower redshift dusty galaxies \citep[][]{Caputi2012,Wiklind2008}. The existence of a large population of undetected galaxies which are not forming stars would not affect the global star formation rate history or inferences about the reionization of the intergalactic medium (IGM), but would affect the estimated cumulative stellar mass as a function of redshift and the number density of passive galaxies at each redshift.

{\bf Acknowledgments} 
JSBW acknowledges the support of an Australian Research Council Laureate Fellowship.  AL was supported in part by NSF grant AST-0907890 and NASA grants
NNX08AL43G and NNA09DB30A. PO acknowledges support by NASA through Hubble Fellowship grant HF-51278.01 awarded by the Space Telescope Science Institute, which is operated by the Association of Universities for Research in Astronomy, Inc., for NASA, under contract NAS 5- 26555.

\newcommand{\noopsort}[1]{}

\bibliographystyle{mn2e}

\bibliography{text}

\label{lastpage}
\end{document}